\documentclass[smallcondensed,final]{svjour3}       
\smartqed  
\usepackage{graphicx}
\usepackage{epstopdf}
\usepackage{amsmath}
\journalname{Journal of Electronic Materials}
\begin{document}

\title{Ab-initio Study of the Electron Mobility in a Functionalized UiO-66 Metal Organic Framework}


\author{Terence D. Musho         \and
        Alhassan S. Yasin 
}


\institute{Terence Musho \at
              Department of Mechanical and Aerospace Engineering
              P.O. Box 6160, Morgantown, WV, 26506-6106 \\
              Tel.: +1-304-293-3256\\
              Fax: +1-304-293-6689\\
              \email{tdmusho@wvu.edu}           
}

\date{Received: date / Accepted: date}

\maketitle

\begin{abstract}
This study leverages density function theory (DFT) accompanied with Boltzmann transport equation approaches to investigate the electronic mobility as a function of inorganic substitution and functionalization in a thermally stable UiO-66 metal organic framework (MOF). The MOFs investigated are based on Zr-UiO-66 MOF with three functionalization groups of benzene dicarboxylate (BDC), BDC functionalized with an amino group (BDC + NH$_2$) and a nitro group (BDC + NO$_2$).  The design space of this study is bound by UiO-66(M)-R, [M=Zr, Ti, Hf; R=BDC, BDC+NO$_2$, BDC+NH$_2$]. The elastic modulus was not found to vary significantly over the structural modification of the design space for either functionalization and inorganic substitution. However, the electron-phonon scattering potential was found to be controllable by up to 30\% through controlled inorganic substitution in the metal clusters of the MOF structure. The highest electron mobility was predicted for a UiO-66(Hf$_5$Zr$_1$) achieving a value of approximately 1.4x10$^{-3}$ cm$^2$/V-s. It was determined that functionalization provides a controlled method of modulating the charge density, while inorganic substitution provides a controlled method of modulating the electronic mobility. Within the proposed design space the electrical conductivity was able to be increased by approximately three times the base conductivity through a combination of inorganic substitution and functionalization.
\end{abstract}

\section{Introduction}
\label{intro}
The ability to tailor the properties of a material for a given task provides a tremendous materials framework that has a range of application. This is especially important in application such as photocatalyst~\cite{yang14,shen15,lee15}, where the material must embody several material attributes to perform tasks efficiently. In the case of a photocatalyst, the important attributes include but are not limited to porosity, photosensitivity, thermal stability, and electronic properties (mobility). A material system that has recently gained more and more interest as a result of the ability to tailor its material attributes are metal-organic frameworks (MOFs)~\cite{canepa13,kui13,shen15,wang15}.  Most noted attributes of MOFs are their higher area density of reaction sites compared to planar catalysts, their high uniqueness of reaction sites, and the ability to easily tailor optical adsorption properties~\cite{tan13,musho15}.  MOF-based structures contain both organic and inorganic portions that make up a crystalline porous network that can be manipulated to achieve different material properties~\cite{shen15,wang15}. For this study, we will limit the design space to focus on a thermally stable UiO-66(Zr)-BDC MOF~\cite{li99} and investigate the influence of different inorganic substitution and functionalization on the electronic mobility. More specifically, this study is interested in understanding how the electronic mobility can be modulated through modification of the material structure. To investigate and predict the electronic mobilities a Density Functional Theory (DFT) approach will be used determine the constitutive properties of a give MOF composition, followed by a Boltzmann based approach to predict electronic mobility~\cite{musho16}.
\begin{figure}[!ht]
\begin{center}
  \includegraphics[width=0.7\textwidth]{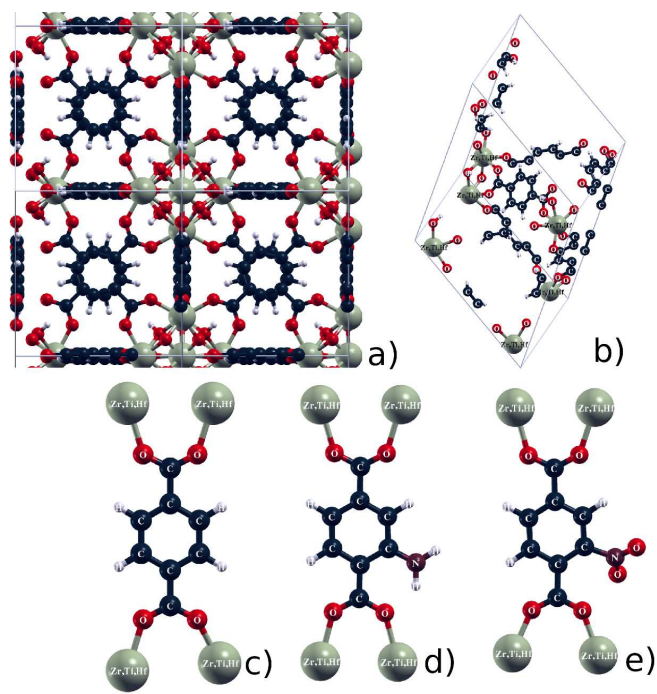}
\caption{Illustration of the UiO-66(M)-R, [M=Zr, Ti, Hf; R=BDC, BDC+NO$_2$, BDC+NH$_2$] unit cell. Sub-figure a illustrates the cubic nature of the structure, b is the 114 atom primitive unit cell, and c-e are the associated functionalized linker designs. }
\label{fig:uc}       
\end{center}
\end{figure}

A difficulty with MOF-based synthesis is their design is \textit{a priori}, this makes the synthesis route extremely important and often a difficult task, coupled with an endless design space which makes material selection a daunting but important task to upgrade photocatalysis device construction~\cite{musho15}.  However, there are several methods of manipulating the structures post-synthesis in the case of a thermally stable UiO-66 based MOF that will allow more mobile charge carriers to be present in the structure.  In fact, studies show that conductivity of MOFs can be altered by five to six orders of magnitude purely based on the implementation of funcationalization and or substitution~\cite{sun15,narayan12,saeki12}.  The first approach, is to use an exchange method to substitutionally exchange the inorganic sites~\cite{musho16}. The inorganic site are sometimes referred to as knots or metalloid clusters are typically situated at the corners of the unit cell. In the case of this study, as pictured in Figure~\ref{fig:uc}, the base structure is comprised of Zr and these Zr atoms are able to be substituted for both Hf and Ti. The second common method of manipulating the electronic properties is to functionalize the organic portion of the MOF. As seen in Figure~\ref{fig:uc}c-e the organic portion of the MOF is used as a linker between the organic knots. These organic linkers are typically comprised of sp2 bonded carbon atoms forming an aromatic terminated carbon chain, in the case of a UiO-BDC MOF with oxygen ions and finally the inorganic knot. This study will focus on three types of functionalization that include a hydrogen group, amino group (NH$_2$, and nitro group (NO$_2$. The photosensitive part of the MOF structure is the aromatic carbon chain this is due to the bonding nature of sp2 bonded carbon atoms. Studies have demonstrated optical absorption properties of MOF-based material to be tailor-able through functionalization from both experimental~\cite{long12} and computational~\cite{musho14,musho15} point of view that allowed band gap modulation of up to 1eV. The origin of the modulation is a result of a change in nature of the sp2 bonding when a functional group is attached to the aromatic carbon ring. Recent studies have gone as far to attribute the amino group attachment as an electron donor and the nitro group attachment as an electron acceptor~\cite{musho14,musho15}.  These findings demonstrate that functionalization results in decreased optical band gap \cite{musho15,musho16} and that the same functionalization may also directly/indirectly influence the mobility of the charge carriers for the given structure~\cite{musho15,kobayashi10,zeng10}.
\begin{table}[h]
\begin{center}
  \begin{tabular}{ l  c  c  c }
  \hline
\hline 
    Linker Design & DFT (eV) & TDDFT (eV) & Experiment UV-VIS (eV)  \\ \hline
    \hline
    UiO-66(Zr)  & 3.10 & 3.78 & 3.76  \\ 
    UiO-66(Zr)-NO$_{2}$ & 2.80 & 2.94 & 2.93 \\ 
    UiO-66(Zr)-NH$_{2}$ & 2.20 & 2.79 & 2.75\\ \hline
    UiO-66(Ti)          &2.67& - &-\\
    UiO-66(Ti)-NO$_{2}$ &2.35& - &-\\
    UiO-66(Ti)-NH$_{2}$ &1.62& - &2.60$^*$\\ \hline
    UiO-66(Hf)          &2.74& - &-\\
    UiO-66(Hf)-NO$_{2}$ &2.66& - &-\\
    UiO-66(Hf)-NH$_{2}$ &1.98& - &-\\
    \hline
  \end{tabular}
  \caption{Summary of calculated and experimentally determined band gap energies for three Zr, Ti, and Hf-based UiO-66 MOF linker designs~\cite{musho16}. Amino Functionalization modulates the band gap by approximately 1eV and inorganic substitution modulates the band gap by another 0.5eV or greater. The asterisks (*) denotes a substituted UiO-66(Zr1Ti5) structure. }
  \label{tab:gap}
\end{center}
\end{table}\\

The attachment of functional groups is a straight forward method of modulating the optical absorption properties. However, the interaction of the functional group with the sp2 carbon turns out to be a complicated interaction. As mentioned in the previous paragraph, the attachment of the functional group introduces an electron donor or acceptor state. The more complicated aspect is that the inorganic knot also plays an important role in the optical properties. By substituting the inorganic ions the band gap can be modulated, not as significantly as functionalization, but less than 1eV. See Table~\ref{tab:gap} for a summary of previously determined values of the band gap. This provides evidence that the inorganic ions influence the bonding of the oxygen atoms and subsequently the aromatic carbon. From visual inspection of the linker coordination, there is directionality in the bonding which indicates a significant amount of covalent interaction between the oxygen-carbon. However, the inorganic-carbon involves more ionic bonding, therefore the scenario arises where the inorganic electrostatically bonds with oxygen which subsequently changes the p-orbital constant of the oxygen and ultimately influencing the hybridization of the sp2 bonds of the aromatic carbon.

From the assessment of the change in band gap, it is reasonable to hypothesize that the electronic mobility can also be manipulated through the aforementioned methods. There have been recent experimental findings that have precisely done this. Several authors have experimentally~\cite{shen15,wang15,hendon13} demonstrated this for a range of MOF designs and determined that the electrical conductivity, which is a function of the electron mobility, can be manipulated considerably. The approach taken in the following study is to use DFT to investigate the origins of the change in mobility as a function of structural change.

For the cases where an MOF embodies intrinsic electron mobility combined with long-range order, as the case of a UiO-66 MOF, the carrier mechanism in these MOF materials is based on two dominate transport modes. The first mode is the mobility of delocalized electrons, which exhibit free electron like conduction (structure has continous states in reciprocal space) or electron tunneling. This is synonymous to the transport found in small length molecular junction. The second mode is a charge hopping mechanism, which is typical for truly organic materials that have extremely localized charges~\cite{tao06}. There is often a trade-off between these two mechanisms. At shorter molecular lengths, the transport is dominated by tunneling and at larger molecular lengths, the transport is governed by electron hopping. The first mechanism is the most interesting as you can take advantage of the delocalized $\pi$ electron in the aromatic carbon ring by modifying their electronic environment through functionalization and inorganic substitution. It is noted by others~\cite{park15} that the charge mobility as result of band type motion can be difficult because the charges are localized and the curvature of the bands are often flat. These flat bands are associated with high effective masses and resemble defect or localized states providing justification for using hopping theory. However, in the case of UiO-66, it has been found in previous studies~\cite{musho15} that the UiO-66 does exhibit band like motion with an effective mass along the short linker design. Therefore, the mobility is hypothesized to be readily modulated by either functionalization or inorganic substitution.\\

Because these materials are a hybrid inorganic composition it is difficult to determine the exact mechanism for the conduction. Typically the mode of electrical conduction in these materials can be broken into several mechanisms. These modes include free electron conduction and electron hopping. Typically in inorganics and ionic conductors, the mode is hopping. It has been determined for the UiO MOF structure this material has semiconducting properties and has free electron conduction along the linkers~\cite{musho15}. By assuming free electron conduction this allows Boltzmann expression based on Bardeen and Shockley's research~\cite{bardeen50} to be applied for UiO MOFs.

\section{Material Design}
The material selected for this study is a UiO-66(M)-R, [M=Zr, Ti, Hf; R=BDC, BDC+NO$_2$, BDC+NH$_2$]. Figure~\ref{fig:uc} is an illustration of the associated MOF design for this study. The organic linker is based on a benzene dicarboxylate (BDC) and is functionalized with and amino group (NH$_{2}$) and a nitro group (NO$_{2}$) resulting in a UiO-66-R, [R = BDC, BDC+NO$_2$, BDC+NH$_2$]. The linker designs are visualized in Figure~\ref{fig:uc}c-e. The UiO-66 structure is a cubic unit cell with an inorganic structure situated at the corners. The 66 designation in UiO-66 denotes a single aromatic ring, where higher numbers designate a greater number of organic ring repetition along the linker. While the porosity can be controlled by increasing the linker length the thermal stability decreases. Similarly, it is important to keep the organic chain short to provide tunneling conductions along the MOF structure. The thermal stability of this MOF has been experimentally determined to be as high as 300-400C. This lends itself well to the application of photocatalyst and thus the selection for this study. While it is beyond the scope of this study to elaborate on the experimental method for inorganic substitution and functionalization, this can be found in the following references~\cite{yang14,lin12,canepa13}.

\section{Computational Method}
A DFT method is used to calculate the ground state properties of the MOF materials. A single primitive unit cell, see Figure~\ref{fig:uc}b, consisting of approximately 114 atoms. Quantum Espresso~\cite{qe} plane wave solver was implemented with pseudized wave functions. The exchange-correlation selected for this study was a Perdew-Burke-Ernzerhof (PBE) functional used with ultrasoft potentials with a cut-off kinetic energy of 680 eV (50 Ry) and a density cut off of 6800 eV (500Ry). Previous studies have investigated the trade-off between BLYP and PBE and their hybrid counterpart and have provided justification for selecting PBE functional~\cite{musho14}. An overview of those findings suggests that the BLYP over-predicts the hydrogen binding resulting in under predicted unit cell volume and band gap. While the hybrid functional can provide a better account of the band gap the introduction of additional empiricism into the prediction was not warranted. 

The k-point mesh was sampled using a Monkhorst-Pack 4x4x4 grid with an offset of 1/4,1/4,1/4. To account for the Van der Waals interactions, a dispersion force correction term~\cite{grimme06,barone09} was incorporated, which introduced some empiricism into the calculation. The scaling parameter (S6) was specified to be 0.75 and cut-off radius for the dispersion interaction was 200 angstroms. Both the ion and unit cell geometries were relaxed to a relative total energy less than 1x10$^{-10}$ and overall cell pressure of less than 0.5kBar. The reader should be made aware that pure DFT predictions of band gap are often under predicted due to the over-analyticity of the functionals and exchange-correlation terms. Therefore, the band gaps reported in this study should not be used as absolutes but used to study the trends.

\subsection{Mobility, Elastic Modulus, and Deformation Potential}
The method used to obtain the mobility was to base this study on previous research~\cite{musho15}, which used a Boltzmann based relaxation time approximation (RTA) approach developed by Bardeen and Shockley for non-polar semiconductors~\cite{bardeen50}. The MOF crystal in this study contains a center of symmetry and therefore is non-polar. In the case of polar crystals, polar optical phone scattering is often a dominating scattering mode and this theory is not appropriate.  The Boltzmann expression developed by Bardeen and Shockley assumes the dominate scattering mechanism is acoustic phonon scattering with electrons. This is deemed the most likely dissipation model for these low modulus and size of the MOF structure, which is well structured for acoustic based phonon modes opposed to optical phonon modes. The phonon-electron scattering is accounted for in the Boltzmann expression by considering a 3D deformation potential. The deformation potential relates the change in the conduction band to a finite amount of strain associated with the acoustic phonon. The expression used to predict the mobility takes the following form,
\begin{equation}
\mu=\frac{q \tau_s}{m^*}=\left(\frac{C_1 \hbar^4}{E_1}\right)^2\left ( m_e m^*\right)^{5/2} (k_B T)^{3/2},
\label{eq:bz}
\end{equation}
here $C_1$ is the modulus, $E_1$ is the three-dimension (3D) deformation potential, $m^*$ is the parabolic effective mass or reduced mass, $m_e$ is the mass of an electron, $T$ is the temperature, and $k_B$ is the Boltzmann constant. The mobility is proportional to the scattering rate divided by the carrier's effective mass. The right portion of Equation~\ref{eq:bz} can be thought of as a closed form expression for estimating the electron-phonon scattering rate ($\tau_s$), which is often the limiting scattering mechanism in single crystal materials. The assumption of phonon limiting scattering mechanism is phonon based dielectric studies of similar MOF materials~\cite{devautour12,warmbier14}. The complex dielectric constant provides a macroscopic viewpoint of the dissipation. A near unity static or real portion of the dielectric constant~\cite{warmbier14} suggests that the charge carriers are highly directional and the Coulomb interactions are minimal. Additionally, nuclear magnetic resonance (NMR) studies~\cite{devautour12} that provide an estimate of the complex portion of the dielectric indicates that there is a noticeable temperature dependence, which is governed by the phonon content.

The effective mass (m$^*$) in Equation~\ref{eq:bz} was determined for the lowest unoccupied molecular orbital (LUMO) in previous studies for the UiO-66 structure~\cite{musho15}. The effective mass relation, $E=\hbar k^2/2m^*$, was fit to both of these minimums, the effective mass is determined to be 8.9m$_e$ for both minimums. It will be assumed for this study that the effective mass is constant for all structures and only the band gap is modulated. It is also assumed that the majority carriers are electrons (n-type conduction).

Additional important parameters in Equation~\ref{eq:bz} are the modulus and acoustic deformation potential. The deformation potential relates a mechanical strain on the unit cell, in this case, a phonon or lattice strain to an electric scattering potential. More specifically, a 3D deformation potential provides a quantity for the probability of an elastic interband scattering of an electron with a phonon that conserves both momentum and energy. There is a trade-off between the mobility of the majority carrier and the phonon or heat production. The other critical parameter is the modulus of the material, which quantifies the resistance of the material to deformation provided a given lattice strain (phonon). Both of these parameters are typically used as fitting parameters within the Boltzmann based mobility calculations. The approach taken in the research is to use first principle approach to determine these parameters.

The modulus of the material, $C_1$, can be related to the bulk modulus of the material. The bulk modulus of the material will govern the sound speed of the material and ultimately the acoustic phonon content. These acoustic phonons are required for interband transitions. To determine the modulus of the material, a series of DFT calculations were conducted that strained the unit cell along the \textless111\textgreater~ direction. Because the charge carriers of the material is not confined within a particular direction and the strain is oriented in all three crystallographic directions, the deformation potential will be 3D deformation potential. 

To determine the modulus, a series of thirteen cases were set-up that range from -1.5\% strain (compression) to +1.5\% strain (tension). The strain is determined by the following, $\epsilon=\Delta l/l_o$, where $l$ is the stained length and $l_o$ is the original length. The strain was applied in all three crystallographic directions and the ion positions are displaced along these directions. The original length ($l_o$) is determined by the relaxed coordinate. A self-consistent field calculation was carried out for thirteen strain positions and the total energy was tabulated. 

The modulus was able to be calculated by relating the total energy to the strain energy. The relationship used to related the volumetric strain energy takes the following form, 
\begin{equation}
E=\frac{1}{2} \bar{V} C_1 (1+\epsilon) \epsilon^2 + E_0,
\label{eq:c}
\end{equation}
where E is the total energy, $ \bar{V}$ is the volume, $C_1$ is the elastic constant, and $\epsilon$ is the strain.
 
It should be pointed out that a difficult aspect of conducting the inorganic substitution study was finding the most stable configuration for low concentration. This is because at low concentrations there are several possible positions for the ion to lie within the primitive unit cell. Referencing the illustration in Figure~\ref{fig:uc}b, there are six possible positions for Zr, Ti, or Hf to occupy. The approach taken in this study follows the procedure of previous investigation~\cite{musho16} and uses a brute force method to find the minimum configuration. This resulting in nearly 729 relaxation trials that had to be executed for each functionalization case. Based on these relaxations, 28 points remain across the compositional space, as will be seen in the following ternary plots in the Results and Discussion Section. While this approach is tractable for three element configurations, anything beyond this becomes too time-consuming. This is an area of current interest where statistical based sublattice methods could improve the throughput of these combinatorial DFT studies.

\section{Results and Discussion}
\begin{figure}[!ht]
\begin{center}
  \includegraphics[width=0.7\textwidth]{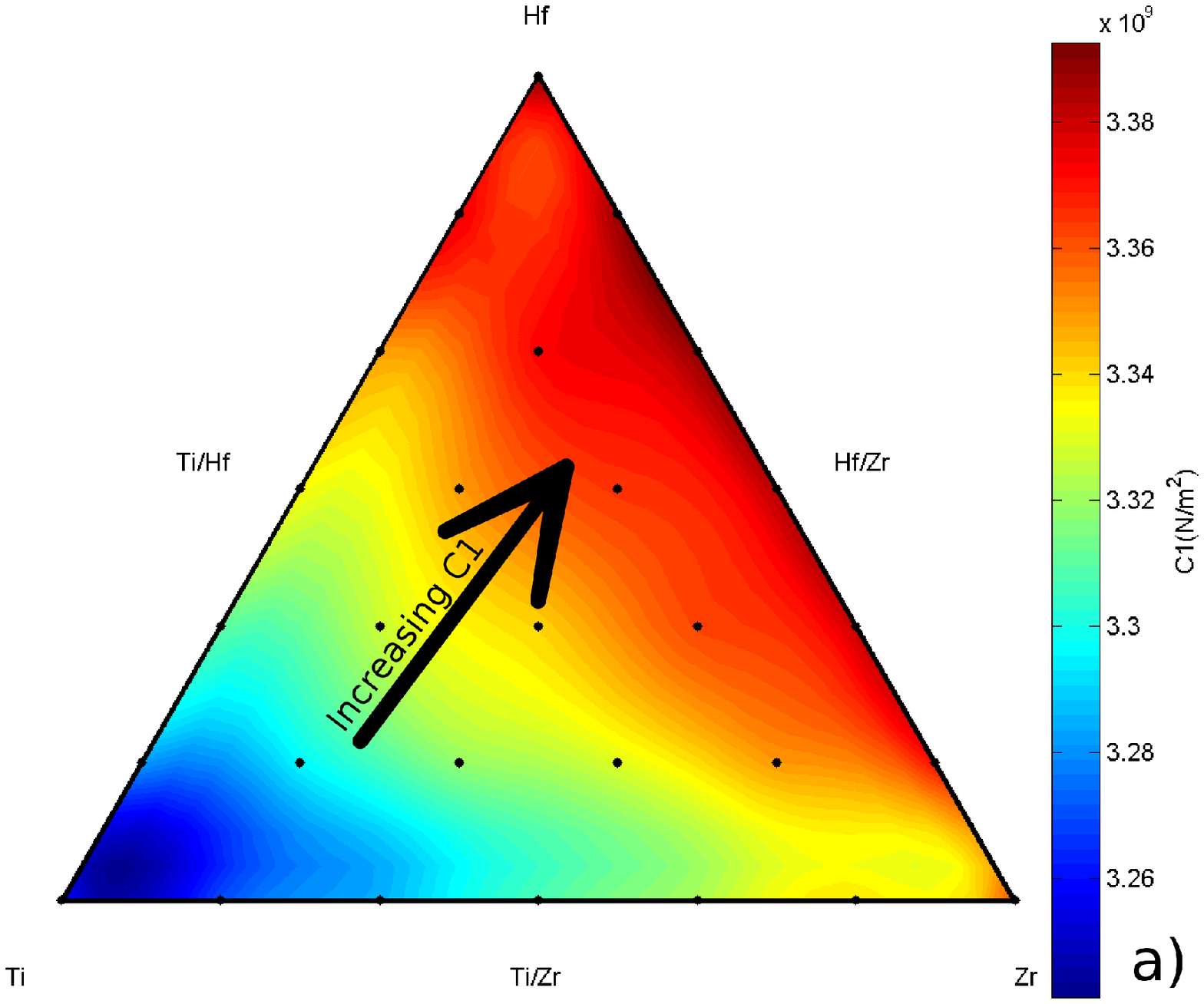}\\
  \includegraphics[width=0.7\textwidth]{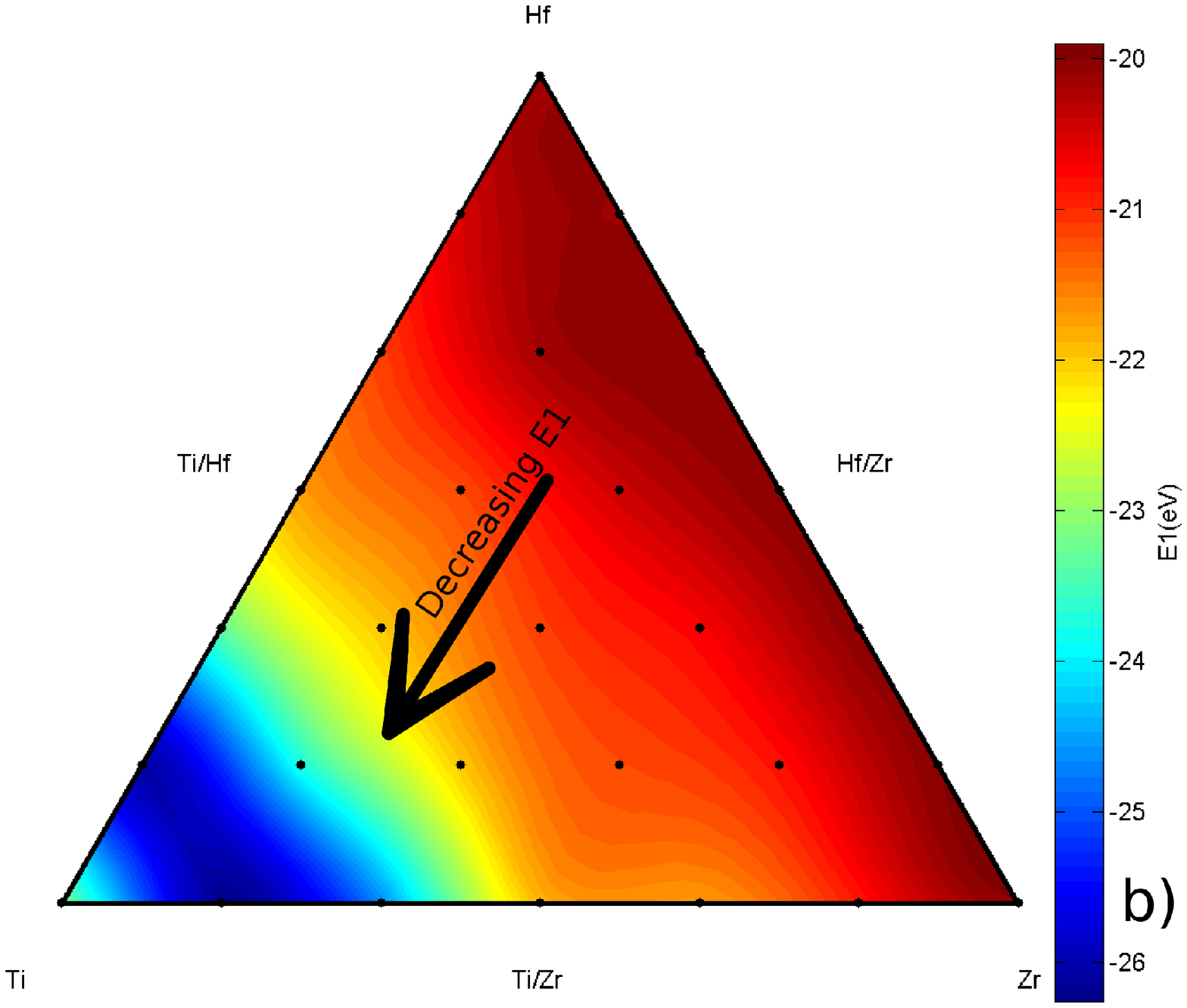}
\caption{Plot of the modulus (C$_1$) and the 3D deformation potential (E$_1$) parameters used in Equation~\ref{eq:bz} to predict the mobility. These results are for the UiO-66(Zr,Ti,Hf)-NH$_2$ MOF design, which has the lowest band gap. The modulus of the material does not change significantly over the space and the deformation potential dominates the mobility. The black dots denote the 28 calculated points and the contour colors are interpolated between these points. }
\label{fig:c1e1}       
\end{center}
\end{figure}
Once the most stable configurations were found using DFT unit cell relaxation method, this configuration was used for the remainder of the DFT calculations that will be discussed in this section. The modeling procedure outlined in the previous section highlighted the parameters necessary to predict the mobility. Using this method the critical parameters were calculated at 28 points throughout the design space that was bound by the three inorganic compositions and the three functionalizations. Figure~\ref{fig:c1e1} is a contour ternary plot of the two critical parameters necessary to calculate the mobility provided by Equation~\ref{eq:bz}. This figure is only for the UiO-66-NH$_2$ functionalization case. The black dots in both of the sub-figures denote the 28 design points that were calculated. The color contours are interpolated between each of these points. Figure~\ref{fig:c1e1}a is the associated elastic modulus parameter, C$_1$. The lowest modulus is associated with the fully substituted titanium based structure with values of 3.26GPa. However, the maximum modulus is between a composition of UiO-66(Hf$_5$Zr$_1$) and UiO-66(Hf$_4$Zr$_2$), with a value of 3.38GPa. The associated range across the entire design space is 0.12GPa, which is no more than 4\% change. This low variability suggests that the modulus is not heavily influenced by the composition of the inorganic knots but rather the compliance of the organic linkers that connect the knots.

Figure~\ref{fig:c1e1}b is the associated 3D deformation potential calculated using the procedure explained in the previous section. The deformation can be thought of as the magnitude of the electron-phonon scattering potential. From Figure~\ref{fig:c1e1}b the trend visualized from the contours suggest the most negative scattering potential is associated with a fully substituted titanium (UiO-66(Ti)) structure. Recall the scattering potential is associated with the amount of LUMO shift for a given amount of strain. Associating this with the modulus from Figure~\ref{fig:c1e1} the material with the lowest modulus has the associated largest absolute value of scattering potential. Referring to the mobility expression in Equation~\ref{eq:bz}, the modulus and scattering potential are inversely related and therefore competing. To maximize mobility you would like to minimize the scattering potential and maximize the modulus. From Figure~\ref{fig:c1e1}b the deformation potential ranges from -20 to -26, which equates to nearly a 30\% variation of the deformation, while the modulus only saw a 4\% variation across the design space.
\begin{figure}[!ht]
\begin{center}
  \includegraphics[width=0.7\textwidth]{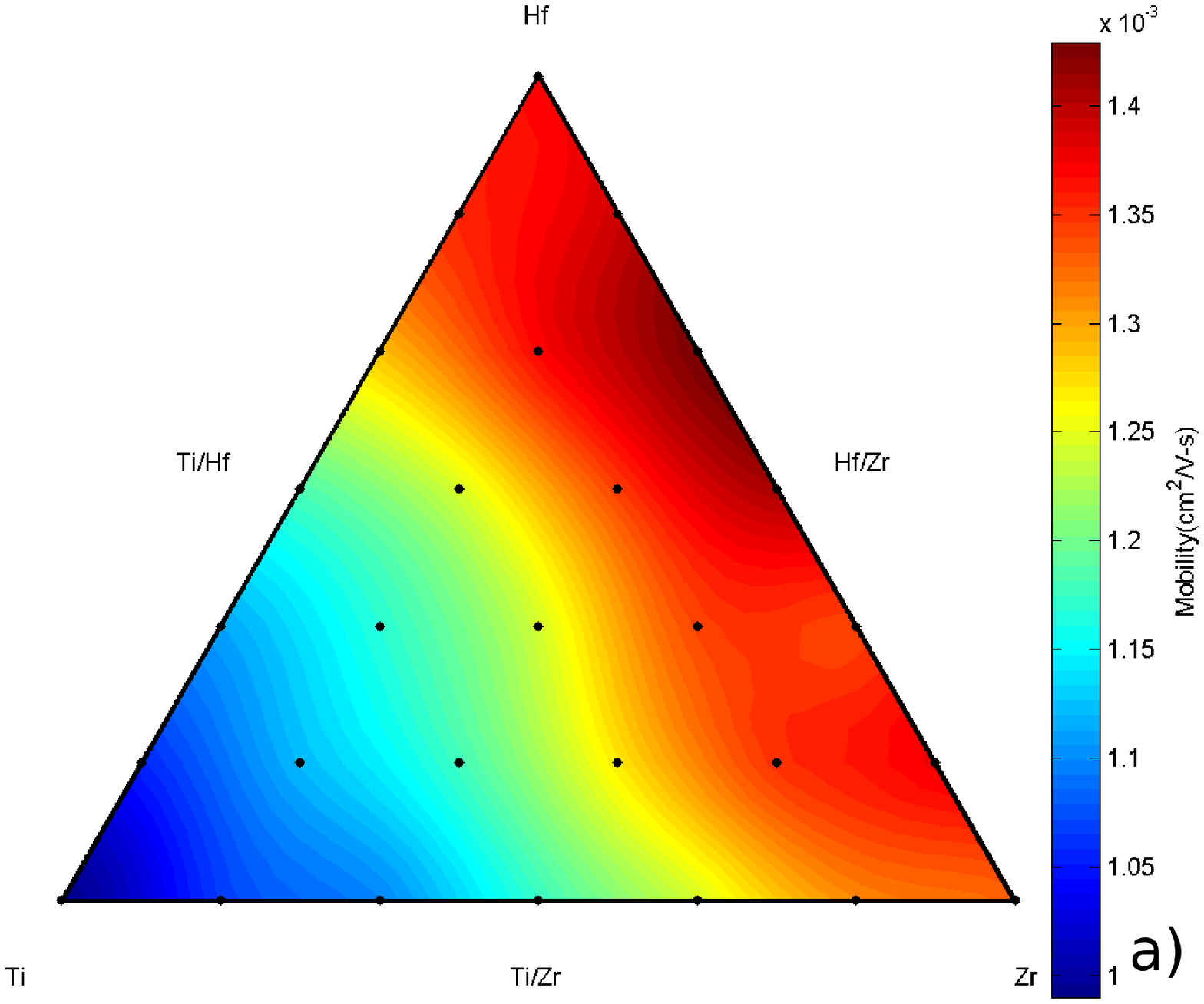}\\
  \includegraphics[width=0.7\textwidth]{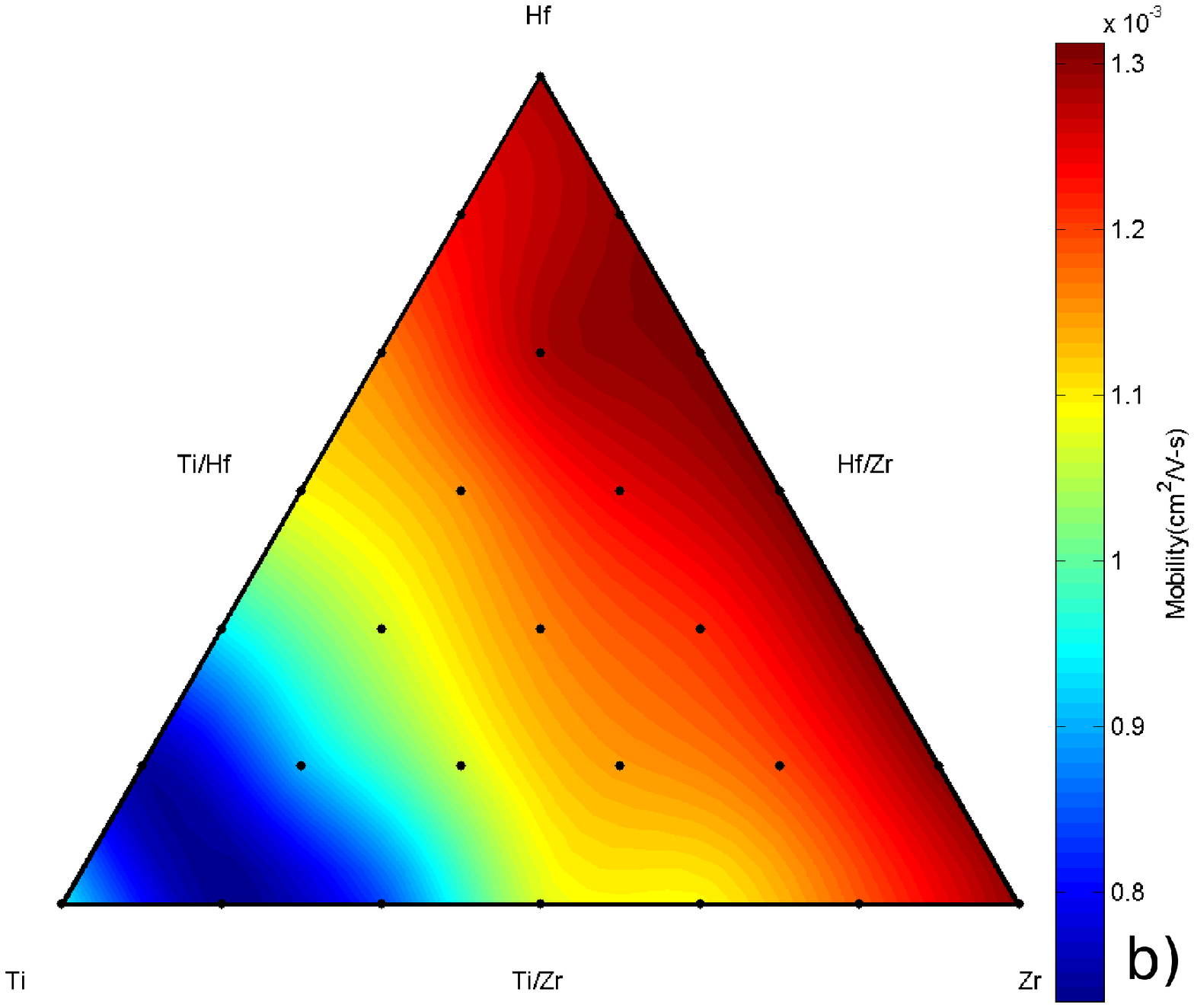}\\
  \includegraphics[width=0.7\textwidth]{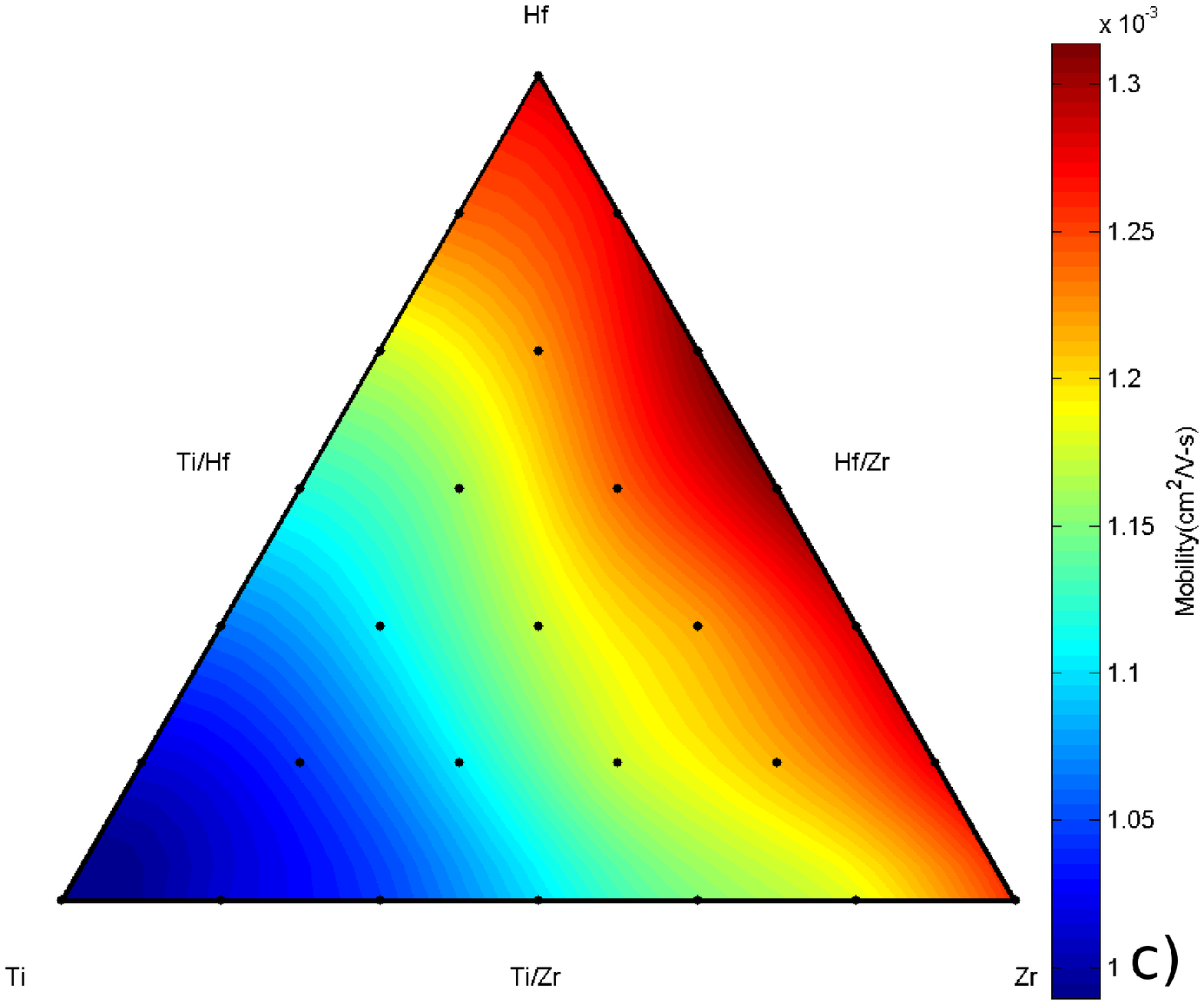}
\caption{Plot of the electron mobility for a range of inorganic knot compositions and organic linker functionalizations. Subfigure a) is UiO66-BDC, b) is UiO66-BDC-NH$_2$, c) UiO66-BDC-NO$_2$. The mobility follows a similar trend as the scattering (deformation) potential provided in Figure~\ref{fig:c1e1}b. The highest mobility is associated with UiO-66(Hf$_5$Zr$_1$)-BDC with an approximate value of 1.4x10$^{-3}$ cm$^2$/V-s.}
\label{fig:mobility}       
\end{center}
\end{figure}

Following the expression in Equation~\ref{eq:bz} and assuming a temperature of 300K and an effective mass of 8.9, an estimate of the mobility was made across the design space for all possible combinations. Figure~\ref{fig:mobility} is the associated mobility predictions. Again the black points represent the DFT calculations and the color contours are interpreted between these points. Each of the subfigures represent a different functionalization case. Across all three cases, there is a very similar trend in the mobility. This trend is also similar to Figure~\ref{fig:c1e1}b for the deformation potential, suggesting that the deformation potential is the dominate term. In the case of MOF design with the largest associated mobility UiO-66(Hf$_5$Zr$_1$)-BDC, is the highest overall with a value of approximately 1.4x10$^{-3}$ cm$^2$/V-s. To put this relative context of other mobility values, the mobility of silicon is 1400 cm$^2$/V-s. This equates to silicon being roughly six orders of magnitude greater mobility. Therefore, the mobility of this MOF is quite low in comparison to an inorganic semiconductor. Looking at the individual ternary plot in Figure~\ref{fig:mobility} the mobility ranges between 0.8x10$^{-3}$ to 1.4x10$^{-3}$ cm$^2$/V. From a percentage perspective, this equates to roughly a 30\% variability in the mobility by substituting the inorganic ions in the MOF lattice. This is not huge compared to inorganic semiconductors but can be considered fairly significant for MOFs.

Focusing on the relative comparison the three ternary plots of Figure ~\ref{fig:mobility} the functionalization does not significantly influence the mobility. There is a slight difference in the contour near UiO-66(Hf$_5$Zr$_1$) but the values do not significantly change. This is an interesting finding because the functionalization significantly modulates the band gap. However, because the functionalization of the organic linker does not influence the compliance or modulus of the overall MOF structure there is not a significant change in the phonon-electron interactions. 

While the mobility is not significantly influenced by the functionalization and only moderately by the inorganic substitution there would be a much larger change in the conductivity. The expression that governs the conductivity takes the following form, $\sigma=q \mu \eta$, where $\sigma$ is the electrical conductivity, $\mu$ is the mobility, and $\eta$ is the electron density. The electron density could be approximated by integrating the Fermi function and the density of states which takes the form of,
\begin{equation}
\begin{split}
{n = \int_{LUMO} g(E)f_D(E)\,dE}	\\
{= \int_{LUMO}\frac{g(E)\,dE}{1 + exp[(E-E_f)/k_BT]},}
\end{split}
\label{eq:n}
\end{equation}
where g(E) is the density of states (DOS), $f_D$(E) is the Fermi-Dirac distribution under the assumption that temperature is at 300K, and $E_f$ is the Fermi energy. 

The DOS values are attained from the results presented in previous studies~\cite{musho14,musho15,musho16} that discussed in-depth inorganic substitution and structure functionalization's affect on the band gap for UiO-66(M)-R, [M=Zr, Ti, Hf; R=BDC, BDC+NO$_2$, BDC+NH$_2$] MOFs.  Fermi level location from DFT calculations suggests that the level for these structures shifted towards the LUMO~\cite{musho16}, this implies that activation energy is thermally dependent and that functional group of BDC+NH$_2$ introduces a donor state and BDC+NO$_2$ introduces an acceptor state, which has been experimentally confirmed~\cite{musho16,sun15,talin14}.

\begin{table}[ht]
\begin{center}
  \begin{tabular}{ l c c c c c }
  \hline
\hline 
Design &C$_1$ (N m$^{-2}$) &E$_1$ (eV) &n (cm$^{-3}$) &$\mu$ (cm$^{2}$ V$^{-1}$ s$^{-1}$) &$\sigma$ (S cm$^{-1}$)  \\ \hline
    \hline
    UiO-66(Zr)  		 &2.9882 x 10$^9$ &-18.654 &1.7960 x 10$^{14}$ &1.3288 x 10$^{-3}$&3.8238 x 10$^{-8}$  \\ 
    UiO-66(Zr)-NO$_{2}$ &3.5920 x 10$^9$ &-21.018 &1.7906 x 10$^{14}$ &1.2583 x 10$^{-3}$&3.6100 x 10$^{-8}$ \\ 
    UiO-66(Zr)-NH$_{2}$ &3.3617 x 10$^9$ &-20.000 &1.9150 x 10$^{14}$ &1.3005 x 10$^{-3}$&3.9904 x 10$^{-8}$ \\ \hline
    UiO-66(Ti)          &2.9033 x 10$^9$ &-21.309 &4.3436 x 10$^{14}$ &9.8948 x 10$^{-4}$&6.8861 x 10$^{-8}$ \\
    UiO-66(Ti)-NO$_{2}$ &3.4729 x 10$^9$ &-23.272 &4.3668 x 10$^{14}$ &9.9230 x 10$^{-4}$&6.9426 x 10$^{-8}$ \\
    UiO-66(Ti)-NH$_{2}$ &3.2590 x 10$^9$ &-23.127 &6.9491 x 10$^{15}$ &9.4293 x 10$^{-4}$&1.0498 x 10$^{-7}$ \\ \hline
    UiO-66(Hf)          &3.0445 x 10$^9$ &-18.509 &1.9145 x 10$^{14}$ &1.3752 x 10$^{-3}$&4.2186 x 10$^{-8}$ \\
    UiO-66(Hf)-NO$_{2}$ &3.6238 x 10$^9$ &-20.872 &2.0109 x 10$^{14}$ &1.2872 x 10$^{-3}$&4.1472 x 10$^{-8}$ \\
    UiO-66(Hf)-NH$_{2}$ &3.3901 x 10$^9$ &-20.181 &2.1896 x 10$^{14}$ &1.2880 x 10$^{-3}$&4.5187 x 10$^{-8}$ \\
    \hline
  \end{tabular}
  \caption{Summary of calculated and elastic constant (C$_1$), deformation potentials (E$_1$), and mobility ($\mu$) for MOF designs at 300K.  The UiO-66(M)-NH$_2$ obtained the lowest mobility that is attributed to the largest deformation potential. The highest conductivity was achieved for a UiO-66(Ti)-NH$_2$ structure.}
  \label{tab:n_m_s}
\end{center}
\end{table}

The location of the band edge will significantly influence the electron density. For smaller band gaps this value will significantly increase resulting in a substantial increase in conductivity.  Table~\ref{tab:n_m_s} shows the calculated electrical conductivity for the MOF compositions, and that the UiO-66(Ti)-NH$_2$ has the highest conductivity even though it has the lowest mobility. This high conductivity is a result of the small band gap that allows for higher concentration of charge carriers. In the case of these materials as presented in Table~\ref{tab:gap} and Table~\ref{tab:n_m_s}, UiO-66(Ti) has the lowest band gap, therefore would have the largest charge density even though this MOF had the lowest mobility. It is noted that BDC-NH$_2$ has the highest concentration due to the smallest band gap attribute for this functional group as seen in Table~\ref{tab:gap}.  The increase carrier concentration is a results of the BDC-NH$_2$ donor nature that influences the LUMO states and results in a smaller band gap. Overall, comparing the lowest conductivity MOF calculated (UiO-66(Zr)-NO$_2$) and the highest (UiO-66(Ti)-NH$_2$) there is nearly a three times increase in conductivity over the design space.  

By being able to to control the mobility and charge concentration allow the conductivity to be controlled for MOF structures using a systematic approach. Functionalization gives the ability to tailor the band gap of MOFs, which allows the influences not only the of light absorption characteristics but also the free electron charge concentration in the structure. Additionally, the composition of the inorganic clusters does influence the band gap and charge concentration but more significantly influences the mobility as a results of modulating the electron-phonon scattering behavior of the structure. This result provides an avenue for using structural modifications of MOF materials to obtain similar trends as moderately doped semiconductors. With the design space of MOF increasing tremendously there are is a requirement to understand how the electrical properties can be modified and potentially increase enough to compete with inorganic semiconductors.

\section{Conclusion}
In this studies a DFT method was used with a relaxation time approximation to predict the electron mobility for three different inorganic knots and three types of functionalization. Findings confirmed that the inorganic substitution can be used to modulate the electron mobility in UiO-66 MOF structures. The deformation potential or the electron-phonon scattering potential can be altered by as much as 30\% over a continuous design space between UiO-66(Zr), UiO-66(Ti), and UiO-66(Hf). The elastic modulus does not significantly change by inorganic substitution or functionalization in a UiO-66 MOF structure. Therefore, the mobility was found to follow a similar trend as the deformation potential with UiO-66(Hf$_5$Zr$_1$) achieving the highest mobility with a value of approximately 1.4x10$^{-3}$ cm$^2$/V-s. While the highest mobility was achieved for a Hf-Zr based MOF the highest electrical conductivity was associated with a Ti because of the increased charge density associated with lowest band gap. Ultimately, this research has identified inorganic subsitituion as a mechanism for controlling the mobility and functionalization a mechanism for controlling conductivity. 

\begin{acknowledgements}
I would like to acknowledge the use of the Super Computing System (Spruce Knob) at WVU, which is funded in part by National Science Foundation EPSCoR Research Infrastructure Improvement Cooperative Agreement \#1003907.
\end{acknowledgements}

\bibliographystyle{spphys}       
\bibliography{journal}   

\begin{thebibliography}{10}
\providecommand{\url}[1]{{#1}}
\providecommand{\urlprefix}{URL }
\expandafter\ifx\csname urlstyle\endcsname\relax
  \providecommand{\doi}[1]{DOI \discretionary{}{}{}#1}\else
  \providecommand{\doi}{DOI \discretionary{}{}{}\begingroup
  \urlstyle{rm}\Url}\fi

\bibitem{yang14}
L.M. Yang, E.~Ganz, S.~Svelle, M.~Tilset, J. Mater. Chem. C \textbf{2}, 7111
  (2014)

\bibitem{shen15}
L.~Shen, R.~Liang, M.~Luo, F.~Jing, L.~Wu, Phys. Chem. Chem. Phys. \textbf{17},
  117 (2015)

\bibitem{lee15}
Y.~Lee, S.~Kim, J.K. Kang, S.M. Cohen, Chem. Commun. \textbf{51}, 5735 (2015)

\bibitem{canepa13}
P.~Canepa, N.~Nijem, Y.J. Chabal, T.~Thonhauser, Phys. Rev. Lett. \textbf{110},
  026102 (2013)

\bibitem{kui13}
K.~Tan, P.~Canepa, Q.~Gong, J.~Liu, D.H. Johnson, A.~Dyevoich, P.K.
  Thallapally, T.~Thonhauser, J.~Li, Y.J. Chabal, Chemistry of Materials
  \textbf{25}(23), 4653 (2013)

\bibitem{wang15}
S.~Wang, X.~Wang, Small \textbf{11}(26), 3097 (2015)

\bibitem{tan13}
K.~Tan, P.~Canepa, Q.~Gong, J.~Liu, D.H. Johnson, A.~Dyevoich, P.K.
  Thallapally, T.~Thonhauser, J.~Li, Y.J. Chabal, Chemistry of Materials
  \textbf{25}(23), 4653 (2013)

\bibitem{musho15}
T.~Musho, N.~Wu, Phys. Chem. Chem. Phys. \textbf{17}, 26160 (2015)

\bibitem{li99}
L.~Hailian, E.~Mohamed, M.~O'Keeffe, O.M. Yaghi, Nature \textbf{402}(6759), 279
  (1999)

\bibitem{musho16}
A.S. Yasin, J.~Li, N.~Wu, T.~Musho, Physical Chemistry Chemical Physics
  \textbf{18}(18), 12748 (2016)

\bibitem{sun15}
L.~Sun, C.H. Hendon, M.A. Minier, A.~Walsh, M.~Dincă, Journal of the American
  Chemical Society \textbf{137}(19), 6164 (2015)

\bibitem{narayan12}
T.C. Narayan, T.~Miyakai, S.~Seki, M.~Dincă, Journal of the American Chemical
  Society \textbf{134}(31), 12932 (2012)

\bibitem{saeki12}
A.~Saeki, Y.~Koizumi, T.~Aida, S.~Seki, Accounts of chemical research
  \textbf{45}(8), 1193 (2012)

\bibitem{long12}
J.~Long, S.~Wang, Z.~Ding, S.~Wang, Y.~Zhou, L.~Huang, X.~Wang, Chem. Commun.
  \textbf{48}, 11656 (2012)

\bibitem{musho14}
T.~Musho, J.~Li, N.~Wu, Phys. Chem. Chem. Phys. \textbf{16}, 23646 (2014).
\newblock \doi{10.1039/C4CP03110E}.
\newblock \urlprefix\url{http://dx.doi.org/10.1039/C4CP03110E}

\bibitem{kobayashi10}
Y.~Kobayashi, B.~Jacobs, M.D. Allendorf, J.R. Long, Chemistry of Materials
  \textbf{22}(14), 4120 (2010)

\bibitem{zeng10}
M.H. Zeng, Q.X. Wang, Y.X. Tan, S.~Hu, H.X. Zhao, L.S. Long, M.~Kurmoo, Journal
  of the American Chemical Society \textbf{132}(8), 2561 (2010)

\bibitem{hendon13}
C.H. Hendon, D.~Tiana, M.~Fontecave, C.~Sanchez, L.~D’arras, C.~Sassoye,
  L.~Rozes, C.~Mellot-Draznieks, A.~Walsh, Journal of the American Chemical
  Society \textbf{135}(30), 10942 (2013)

\bibitem{tao06}
N.J. Tao, Nat Nano \textbf{1}(3), 173 (2006)

\bibitem{park15}
S.S. Park, E.R. Hontz, L.~Sun, C.H. Hendon, A.~Walsh, T.~Van~Voorhis,
  M.~Dincă, Journal of the American Chemical Society \textbf{137}(5), 1774
  (2015)

\bibitem{bardeen50}
J.~Bardeen, W.~Shockley, Phys. Rev. \textbf{80}, 72 (1950).
\newblock \doi{10.1103/PhysRev.80.72}.
\newblock \urlprefix\url{http://link.aps.org/doi/10.1103/PhysRev.80.72}

\bibitem{lin12}
C.K. Lin, D.~Zhao, W.Y. Gao, Z.~Yang, J.~Ye, T.~Xu, Q.~Ge, S.~Ma, D.J. Liu,
  Inorganic Chemistry \textbf{51}(16), 9039 (2012)

\bibitem{qe}
P.~Giannozzi, S.~Baroni, N.~Bonini, M.~Calandra, R.~Car, C.~Cavazzoni,
  D.~Ceresoli, G.L. Chiarotti, M.~Cococcioni, I.~Dabo, A.D. Corso,
  S.~de~Gironcoli, S.~Fabris, G.~Fratesi, R.~Gebauer, U.~Gerstmann,
  C.~Gougoussis, A.~Kokalj, M.~Lazzeri, L.~Martin-Samos, N.~Marzari, F.~Mauri,
  R.~Mazzarello, S.~Paolini, A.~Pasquarello, L.~Paulatto, C.~Sbraccia,
  S.~Scandolo, G.~Sclauzero, A.~Seitsonen, A.~Smogunov, P.~Umari, R.M.
  Wentzcovitch, Journal of Physics: Condensed Matter \textbf{21}(39), 395502
  (2009)

\bibitem{grimme06}
S.~Grimme, Journal of Computational Chemistry \textbf{27}(15), 1787 (2006).
\newblock \doi{10.1002/jcc.20495}.
\newblock \urlprefix\url{http://dx.doi.org/10.1002/jcc.20495}

\bibitem{barone09}
V.~Barone, M.~Casarin, D.~Forrer, M.~Pavone, M.~Sambi, A.~Vittadini, Journal of
  Computational Chemistry \textbf{30}(6), 934 (2009).
\newblock \doi{10.1002/jcc.21112}.
\newblock \urlprefix\url{http://dx.doi.org/10.1002/jcc.21112}

\bibitem{devautour12}
S.~Devautour-Vinot, G.~Maurin, C.~Serre, P.~Horcajada, D.~Paula~da Cunha,
  V.~Guillerm, E.~de~Souza~Costa, F.~Taulelle, C.~Martineau, Chemistry of
  Materials \textbf{24}(11), 2168 (2012)

\bibitem{warmbier14}
R.~Warmbier, A.~Quandt, G.~Seifert, The Journal of Physical Chemistry C
  \textbf{118}(22), 11799 (2014)

\bibitem{talin14}
A.A. Talin, A.~Centrone, A.C. Ford, M.E. Foster, V.~Stavila, P.~Haney, R.A.
  Kinney, V.~Szalai, F.~El~Gabaly, H.P. Yoon, et~al., Science
  \textbf{343}(6166), 66 (2014)

\end{thebibliography}

%
%

\end{document}